# Structure and mechanical characterization of DNA *i–motif* nanowires by molecular dynamics simulation

**Raghvendra Pratap Singh**[1,2], **Ralf Blossey**[2] **and Fabrizio Cleri**[2,*]

[1] *Institut d'Electronique Microelectronique et Nanotechnologie (IEMN UMR Cnrs 8520), University of Lille I, 59652 Villeneuve d'Ascq, France*

[2] *Interdisciplinary Research Institute (IRI USR Cnrs 3078), University of Lille I, 59658 Villeneuve d'Ascq, France*

Running title: **Structure and mechanics of DNA *i–motif***

(*) Corresponding author: fabrizio.cleri@univ-lille1.fr, tel: +39 320 197928

## ABSTRACT

We studied the structure and mechanical properties of DNA *i–motif* nanowires by means of molecular dynamics computer simulations. We built up to 230 nm-long nanowires, based on a repeated $TC_5$ sequence from crystallographic data, fully relaxed and equilibrated in water. The unusual C•$C^+$ stacked structure, formed by four *ss*DNA strands arranged in an intercalated tetramer, is here fully characterized both statically and dynamically. By applying stretching, compression and bending deformations with the steered molecular dynamics and umbrella sampling methods, we extract the apparent Young's and bending moduli of the nanowire, as well as estimates for the tensile strength and persistence length. According to our results, the *i–motif* nanowire shares similarities with structural proteins, as far as its tensile stiffness, but is closer to nucleic acids and flexible proteins, as far as its bending rigidity is concerned. Furthermore, thanks to its very thin cross section, the apparent tensile toughness is close to that of a metal. Besides their yet to be clarified biological significance, *i–motif* nanowires may qualify as interesting candidates for nanotechnology templates, due to such outstanding mechanical properties.

## INTRODUCTION

Among the non–standard DNA structures constructed from base pairs not following the standard Watson–Crick (WC) association rule, the *i–motif* is the most recently identified (1-3). Its existence has been assessed *in vitro*, under acidic pH conditions, as a tetrameric structure formed by four intercalated DNA strands, held together by protonated cytosine–cytosine, or $C \bullet C^+$, pairs. However, *i–motif* tetramers as well as G–tetraplex have also been observed *in vivo* (4,5), most notably in the terminal part of the human genes, or telomere, where rather long (50–210 bases) asymmetric G–rich and C–rich single–stranded portions of DNA are found. At present, it is not yet clear whether such DNA tetraplex structures are stably formed and used by eukaryotes, while the duplex–tetraplex interconversion has been studied in some detail (3). From such studies it was observed that at physiologic conditions of pH, salt concentration and temperature, the C–rich and G–rich strands form a normal Watson–Crick duplex, whereas at pH between 4.5 and 5 the *i–motif* and the G–quadruplex become the most stable and abundant forms of supramolecular association.

The stabilization of poly–C DNA strands upon lowering the pH, and the resulting $C \bullet C$ pairing, has been attributed to the hemi–protonation of the C bases (6,7). One extra hydrogen bond is formed when a proton is resonantly exchanged between the two facing N atoms of the base pair (Fig. 1a), now a $C \bullet C^+$, which leads to stabilization of such a non–WC structure. Hemi– (or resonant) protonation of C bases is clearly characterized by two Raman lines at 1385 and 1542 $cm^{-1}$ (8), while the resulting tetrameric arrangement of the sugar–phospate backbone (Fig. 1b) with its unique stacking, can be recognized by the fingerprints at 804, 852, 888, and 972 $cm^{-1}$ (8).

The identification of the elementary process of resonant protonation above, led to the suggestion of possible *i–motif* formation pathways, which were experimentally tested on extended C–rich nanowire tetrameric structures formed by $C_7$ (8), $C_7GC_4$ (9), and $TC_n$ ($n$=3,4,5) (10) base motifs. A notable information that emerges from such studies is that the formation of the tetrameric stacked structure should not proceed through association of preformed dimers, but rather by fast, successive intercalation of single strands into short–lived dimer and trimer species. The fact that $C \bullet C$ dimers are not observed to compete with the tetramers at low enough pH shows that the C–rich dimer is fairly unstable on the time scale of the tetramer association reaction. In the following we will demonstrate by our molecular simulations that this is indeed the case.

The stability of the tetrameric nanowire structure is found to increase rapidly with the number of $C \bullet C^+$ pairings per unit, as shown in (10), the lifetime (in minutes) of $TC_n$ tetramers, obtained by NMR NOESY and TOCSY spectra, increasing by about two orders of magnitude for each unitary increase of $n$.

From a theoretical point of view, atomistic simulations (11–13) indicate that the stabilization energy of the $C \bullet C^+$ pairing indeed results from a subtle balance between the increased electrostatic repulsion of the $PO_4$ groups, closer in the tetramer than in standard *ds*DNA; favorable dipolar interactions between the C–N and C–O dipoles in stacked pairs; and the extra dipolar interaction plus H–bond energy, induced by the resonant protonation of N$^{...}$N atoms of the facing cytosines. Recent molecular dynamics (MD) simulations (14) confirmed the stacking structure, and underscored the role of sugar–sugar interactions along the backbone as a further stabilizing element.

Besides their possible role in the genome, still awaiting a full clarification, such DNA nanowires can be also attractive in the domain of bio–inspired materials for nanotechnologies.

Notably, various kinds of biomimetic nanowires have been already obtained from B–DNA (15), proteins (16), and even from viral particles (17). Electrical, optical, plasmonic features have been added to such wires by metallization, wherein metals have been "coated" or "moulded" onto the outer or inner surfaces of these biomolecular templates (18-20). The *i–motif* could as well be a good candidate for nano-templating, being easily manipulable and apparently stable over quite long time scales (8). However, while its structure is rather well assessed, a thorough mechanical characterization of such bio–nanowire is still lacking, and even such basic quantities as the persistence length have not been described yet. One of the main results of the present study will be to shed light on such important mechanical features.

In the present work, we studied the mechanical properties of *i–motif* nanowires by means of all–atom MD computer simulations. We built very long tetramers, up to 230 nm, based on the $TC_5$ base unit embedded in a shell of water with appropriate counter ions. We made use of several computational techniques based on molecular constraints, as well as the "steered– MD" (21,22) and "umbrella sampling" (23) methods, to simulate the application of external forces to the *i–motif*. According to the eigenmode of deformation, we extract the equivalent Young's modulus, toughness, bending stiffness, and hence the persistence length of the nanowire. We compare our results to the corresponding data for *ds*DNA with the same base sequence, and to several other biopolymers. Finally, some biological and nanotechnology implications of our findings are discussed.

## MATERIALS AND METHODS

We constructed the atomistic model of the *i–motif* from the NMR crystallographic structure of $(TC_5)_4$ tetramer, solved by Gehring, Leroy and Guéron (1) (PDB id: 225D). Hydrogen atoms were added to the topology following the *Charmm27* force–field guess method. The extra proton on N3 of cytosine was added, with initial orientation adjusted according to the experimental indications (9).

According to the suggestions of Leroy et al. (2,9), protonation of the N3 atom of cytosine, necessary for the formation and stability of this supra–molecular structure could be realized in a few different ways: (i) cytosines belonging to adjoining strands are alternately protonated along the ladder; (ii) two out of four strands are completely protonated at N3 of cytosine, while the two others carry the usual cytosine nucleotide; (iii) the charges (proton of N3) are situated around, and separated by, the narrow groove; (iv) the charges are situated around, and separated by, the wide groove of the *i–motif*. From all of these possibilities we have chosen the second one, namely to start from a structure with two nearby chains fully protonated, while the other two were normal cytosine nucleotides. This choice was motivated because we wanted to check the possibility of protons dynamically jumping from one chain to the other, as well as the implementation of modified force–field parameters. In fact, it turned out to be easier to implement a new set of parameters for just two of the chains, while using the normal force–field parameters for the other two, rather than a new set of hybrid parameters for all the four half–protonated hybrid chains.

Energy minimization of the isolated structure at constant volume was firstly carried out with the NAMD software (21). The structural energy of the system was again minimized after diluting the molecule in water, where the maximum compactness was attained, and an extra hydrogen bond was formed between the C•C pairs. We used the TIP3P water model in all of our simulations. The hydrated structure of the fragment with a length $L$=3.6 nm, was subsequently used as the basic unit for all the foregoing input structures and simulations.

Initial structures for longer fragments $n[(TC_5)_4]$ were constructed by translating $n$=4, 8, 16, 32 or 64 times the basic unit coordinates along the z–axis of the optimized geometry of the relaxed $(TC_5)_4$ unit. Chimera (25) and VMD (26) molecular modeling packages were extensively used in this stage, also to generate the missing terminal residues (P, O1P and O2P), which are the essential backbone atoms participating in the formation of the phosphodiester bonds to join the prime (3' and 5') ends of the previous nucleotide with the next in-line along the backbone. In this way, we generated *i–motif* structures up to about 230 nm length for the $32[(TC_5)_4]$.

We performed finite temperature MD simulations at T=300K, of: *i–motif* DNA tetramers of different lengths, both protonated and non–protonated; a dimer with 8 basic units ($L$=28.8 nm) and the same sequence, i.e., $8[(TC_5)_2]$ (akin to a *ds*DNA with a 8x(dTC_5) sequence); and on a single strand of *i–motif* monomer of the same length, i.e. $8[(TC_5)]$ (akin to a *ss*DNA with a 8x(TC_5) sequence). For each of these systems, we initially performed a fully restrained minimization (i.e., with the DNA fragments immobile) in a large box of TIP3P water with XYZ size of 9x9x($L$+6) nm³, and ions at 0.2 mM/liter, to let water and ions settle at their optimum starting positions. Subsequently, a position–restrained minimization of the system was performed, with restraints still applied on the phosphodiester quadruple backbone of the *i–motif*. Eventually, we ran all–atom unconstrained minimization under constant–{NPT} conditions, followed by quenching to obtain the best starting structures with the lowest energy at T=0 K. Finite–temperature equilibration runs at T=300 K and constant–{NVT} were subsequently performed for 2 ns, followed by several "production runs" of MD simulations of length of 5 up to 10 ns, saving the atomic coordinates every 1 ps.

Steered molecular dynamics (SMD) simulations were performed, to study the elasticity under stretching and bending of the hydrated *i–motif* structures, using the constant–velocity or constant–force protocols of NAMD. To perform SMD, we kept fixed either 4 atoms (for *i–motif* tetramers), or 2 atoms (for *i–motif* dimers) at one end of the structure. Similarly, at the opposite end, 4 or 2 atoms, called SMD–atoms, were selected to carry the applied constant velocity. Under the applied perturbation to the SMD–atoms at one end, while holding still the fixed–atoms at the other end, the whole *i–motif* structure responds by developing a steady deformation, tension, compression, bending, according to the vector of the applied perturbation.

Such SMD simulations were performed for all of the four–, double–, and single–stranded *i–motif* structures described above. The center of pull was calculated by averaging the coordinates of all the SMD atoms. The direction of pull was identified by the direction of the vector connecting the fixed and SMD atoms. In practice, the center of pull of the SMD atoms was attached to a dummy atom by a virtual spring (21,22). The constant velocity along the z–coordinate axis was applied to the dummy atom and the force was measured between dummy and SMD atoms. After several tests, a virtual spring constant of 1 kcal mol⁻¹ Å⁻² was adopted (1 kcal mol⁻¹ Å⁻² = 695 pN nm⁻¹) together with typical pulling speeds of 1 to 3 ms⁻¹, to ensure a reasonable signal–to–noise ratio.

Bending was also simulated by means of the "umbrella sampling" technique (23). We used a harmonic biasing potential, only to simulate the action of a force at the midpoint of the nanowire (the central nucleotide tetrad, one base per chain), while holding still the two extremes. In this way we could drive the nanostructure along the bending deformation trajectory, with a pulling velocity of 5 ms⁻¹ up to reaching the maximum deformation of 5 nm. The umbrella potential provided the confinement in the required deformed structure, while further relaxation steps were performed, to explore the most stable conformations and to extract the corresponding values of deformation energy and force along the trajectory.

## RESULTS

### Molecular structure characterization

Since we started by rigidly duplicating a small chunk of *i–motif* tetramer, namely a tetrameric fragment obtained from the x–ray crystallographic structure, we found it necessary to compare the equilibrated structures of all our *i–motif* systems, both protonated and non–protonated, as well as the dimer with the same sequence, to gain a better insight about the structural features and dynamical stability of our systems. In Figure 1 we show: (a) the structure of the relaxed, protonated dimer, with a repeated $[(TC_5)_2]$ (or equivalently, $d(TC_5)$) structure; (b) a side–view of the relaxed, protonated *i–motif* tetramer, $[(TC_5)_4]$, formed by intercalation of two identical $[(TC_5)_2]$ dimers; and (c) the top–view of the same tetramer.

From the crystallographic coordinates of the *i–motif* basic unit, the average interplanar distance between adjacent cytosines along a same chain is 0.62 nm, while the intercalating distance between adjacent cytosines belonging to different chains is 0.31 nm (see Figure 2 for a definition of such distances). Therefore, the nucleotides in each single chain (out of four in the *i–motif*) have larger distances than their classical B–DNA counterpart. This agrees with the difference in contour length $L$, between the *i–motif* and a B–DNA with the same nucleotide number and sequence (in fact, 50 bp correspond to $L=16.3$ nm in B–DNA, and $L=28.85$ nm in the *i–motif*). After the full–energy optimization, the T=0 K relaxed *i–motif* structures attained an average interplanar distance of 0.64 nm between the nucleotides of same chain, or base–pairs of parallel pair of chains (the dimer), whereas the average intercalating distance settled to 0.32 nm, close to the stable value of B–DNA. The two intercalating dimers are anti–parallel to each other, arranged like the 5'-3'||5'-3' and 3'-5'||3'-5' ordinary DNA duplex.

In B–DNA with random sequence, the interplanar distance on both strands is of about 0.34 nm, while in each of the two dimers making up the *i–motif* it is about 0.64 nm. This means that each dimer composing the *i–motif* is much less twisted than B-DNA: indeed, we measure an average twist angle of ~12° compared to ~34° in B-DNA, with a correspondingly higher helical rise per base pair. However, it should be noted that in the experiments (1,7,9) dimers are very short lived, and either decay back to monomers, or directly evolve to tetramers, by some yet to be clarified mechanism.

It is also worth noting that the *i–motif* structure displays a "minor" and a "major" groove (see Fig. 1). However, this nomenclature must not be confused with the corresponding features of B–DNA, in which the tertiary structure (the helical folding) gives rise to two grooves. In the *i–motif*, the two grooves are both due to the secondary structure of the strands, whose intercalated pairs are rotated by ~50° about the common z–axis (Fig. 1c), thus defining two sets of average distances between the P backbones, (Fig. 2a).

The large interplanar distance in the dimer which, as said, means a larger base–pair rise, as well as smaller tilting angle, sliding and bending parameters per base–pair, should make in principle the *i–motif* structure stiffer and more straight compared to B-DNA. The average diameter is $D=1.85\pm0.05$ nm, calculated by assuming an average cylindrical shape delimited by the backbone P atoms of the four chains. It is worth noting that such a less twisted, more straight structure makes the *i–motif* also thinner than B–DNA, despite being formed by twice as many nucleotides, with a resulting nearly doubled atomic density. The more detailed structure analysis shows that the diameter differs between cytosine–rich regions and intercalated thymine tetrads. An inner cylindrical region can be also identified, based on the average distance between N3 atoms of facing nucleotides from the four chains, whose diameter is $D_i=0.35 \pm 0.03$ nm.

## *Dynamical stability*

The sugar pucker of the ribose ring is an important feature indicating the helical shape of the backbone. In nucleic acids the sugar pucker is responsible for their helical nature. The five–membered ring of the ribose sugar cannot be planar because of steric reasons, therefore one or two atoms move typically by 50 pm above (*endo*) or below (*exo*) the plane of the sugar ring. In DNA, the C2'–endo and C3'–endo conformations are found to be in equilibrium, and help DNA to acquire its normal B–form, while a predominance of –exo conformations would rather push the DNA to acquire its hydrated A–form. On the other hand, RNA molecules strictly follow C3'–endo sugar pucker.

In our thermally equilibrated *i–motif* structures we always observed the C3'–endo configuration (i.e., the carbon C3' moves above the plane C1'–O4'–C4') for all the cytosine pairs, including the thymine nucleotides at the 5' terminus, except the 3'-terminal cytosine, which was found to switch dynamically between C4'–exo and C3'–exo sugar pucker. We also observed a temporary switching to the C4'–endo for some nucleotides, a relatively rare event during the course of very long MD trajectories.

The next important indicator in the stabilization of the *i–motif* structure is the conformation of thymine nucleotide tetrad (i.e., the layer where the four T's from the four intercalating chains meet). Experimental studies show that in this tetrad of thymines, one of the pairs is always hydrogen-bound and paired, while the second pair could either bind inward, or extend outwards from the cylindrical core structure (28). Interestingly, after MD relaxation we found some thymine tetrads (i.e., a layer of T bases, separated by five C tetrads, in the repeated (TC$_5$) four-stranded structure) behaving exactly in this way. Typically, we observed one thymine tetrad with an extended outwards pair (see arrows in the next Fig. 3a), alternating to a nearby tetrad with all the four T's inward bound and properly paired.

In order to demonstrate the stabilization of the structure ensured by the protonation, we compared the dynamical evolution of the fully protonated 8[(TC$_5$)$_4$] *i–motif* and the non-protonated (*np*) tetramer with the same nucleotide composition, 8[(dTC$_5$)$_2$] , over a time scale of 5 ns (in practical terms, the two starting structure were identical, apart from the extra proton). In Figure 3 we show a comparison of the structures of the *i–motif* (a) and the *np*–tetramer (c), starting from the same initial 8-unit configuration (b). We find that during this time the *np*–tetramer crumbles, and looses its *i–motif*–like initial structure, becoming a rather disordered structure with different interplanar distances for every base pair. In Figure 4 we also show, for better clarity, only one chain each for the *i–motif* (a) and the *np*–tetramer (b), in each case comparing the instantaneous conformation of the chain after 5 ns of MD, against the initial structure. For the *np*–tetramer, the backbone turns from a nearly straight wire to a somewhat helicoidal form, while the core of the structure is destroyed. By contrast, the protonated *i–motif* maintains a linear and straight backbone, with constant interplanar and intercalation distances, eventually adjusting to an even more straight conformation and reorganizing the initial defects introduced by the artificial periodic construction. All this findings clearly support the idea of the importance of protonation, in establishing the form and integrity of the *i–motif* structure.

Since the two identical *n*[(TC$_5$)$_2$] dimers are intercalated, each T•T and C•C+ pair of hydrogen bonded nucleotides (at average distance of 0.62 nm) from one dimer, will have either a T•T or a C•C+ pair from the other dimer, as a nearest neighbor (at the average distance of 0.32 nm). Therefore, in our *i–motif* tetramer there are three different set of tetrads, formed by adjacent base–pairs from different dimers: a purely C•C+ tetrad; a purely T•T tetrad; and a mixed C•C+/T•T tetrad. Whilst the overall behavior of the structure is conserved during the dynamics, these tetrads follow however a slightly different behavior in their respective local environment. The C•C+ tetrad shows a high degree of compactness, and the distances between inter- and intra-backbone P atoms are within the experimental ranges. For the *i–motif*, the inter-strand P distances is 0.79 nm (NMR value 0.76 ± 0.03 nm (27)) and intra-strand P distance is 1.53 nm (NMR value 1.54 ± 0.03 nm (27)). The behavior of the T•T tetrad is different, in that only one pair of thymine nucleotides is allowed to be in pairing by using its two hydrogen bonds, and stays within the cylindrical *i-motif* geometry, while the other two T's of the tetrad can move away leaning outwards from the core structure (see again Fig. 3a). Such a displacement forces also the neighboring cytosines (which are also part of the mixed C•C+/T•T tetrad) to deviate from the typical pattern followed by the C•C+ pairs throughout the length of the *i–motif*. Notably, this results in larger P–P distances across the backbone, and a locally wider groove architecture (24). Overall, both the C•C+/T•T and T•T regions are a little wider on average than the C•C+ ones.

### Water and ions distribution

Positioning of the water molecules in close-by solvation shells is another very important factor in stability and functionality of nucleic acids in their cellular environment. Hence we also studied the water and ion concentration in close-by solvation shells for our *i–motif* structures. The average density of water molecules, and ions per unit volume, are shown in Figure 5a and 5b, respectively; a typical cross section for a thickness of 6 nm along z is shown in Fig. 5c. We find that water molecules in close shells tend to penetrate the major groove (green region in Fig. 5c), while they align along, but not inside, the minor groove of the *i–motif* (orange region). Such an apparent hydrophobic behavior of the minor groove limits the solvent accessibility (2), which in fact attains the normal STP water density only at about 0.6 nm away from the outer surface of the nanowire (Fig. 5a). This effect reduces the backbone repulsion, with an overall stabilizing effect on the *i–motif* structure. It is worth noting that in the case of B–DNA, water is known to have cooperative effects on binding of ligands and DNA binding proteins (28), while in case of the *i–motif* the minor groove shows a rather high degree of hydrophobicity.

On the other hand, the radial distribution of counter ions about the *i–motif* phosphate backbone in Fig. 5b is in qualitative agreement with the shape of the radial distributions generally observed for B–DNA (29). Notably, $Na^+$ ions tend to cluster around the $PO_4^-$ backbone groups, in this case mostly around the (more hydrophobic) minor groove with a noticeable subsurface peak of ~0.3 mM/l (see also Fig. 5c). Around the nanowire we find a concentration of ~0.7 mM/l, within a distance of 1 nm from the surface. Conversely, $Ca^-$ ions are nearly absent up to a large distance from the surface, with just a small peak at 1 nm at a concentration about 5 times smaller than $Na^+$; a similar effect is observed in B–DNA, however with a larger ratio of about 10 between positive and negative counter ions (29). After ~1.5 nm, the $Ca^-$ concentration settles to the same constant value of 0.17 mM/l, similar to $Na^+$ at the same distance, ensuring a neutral solution on average.

### Uniaxial stretching modulus

The elastic behavior of biomolecules under uniaxial deformation has been studied experimentally by a variety of techniques, such as optical or magnetic tweezers (see e.g. (30)), or atomic force microscopy (AFM, see e.g. (31) and references therein). After a number of phenomenological studies mainly aiming at the interpretation of the experimental data, the theory behind such experiments is by now pretty well understood (see e.g. (32–34)). In a series of recent papers, the correspondence between different experimental conditions, and polymers with widely different molecular structures, was framed within a robust statistical mechanics framework (35–37).

   Molecular simulations have been extensively and successfully used to estimate the elastic moduli of various biomolecules, covering a wide spectrum of situations (see (38) and references therein). In order to bypass the entropic elasticity regime, we started from a fully extended nanowire along the axial direction, with contour length $L$=28.8 nm. Next, we applied the tensile deformation along the nanowire axis (see Materials and Methods). Since we had no prior knowledge about the possible range of values of the stretching modulus, the most critical step in such a computer experiment was to choose the right combination of pulling velocity and spring constant for the "spacer" spring (between the set of dummy atoms and the pull group). The main problem is that, for linear elasticity to be valid, one has to remain in a regime of extremely small deformations, which, in a complex molecular structure, makes for a very noisy estimate of a fluctuating force at fixed small displacements. We tried various combinations of spring constant and pull velocity, and finally we settled upon a spring constant of 1 kcal mol$^{-1}$ Å$^{-2}$ or ~0.7 Nm$^{-1}$ (comparable to an AFM tip) and pulling velocity of 1 ms$^{-1}$ (in fact, much faster than any experiment), which ensured a reasonably slow and steady response of the nanowire.

   Figure 6 shows the typical result of a uniaxial deformation experiment, both in compression and in tension. The force-displacement plot under such conditions is indeed quite noisy, and displays a moderately oscillatory shape due to the non-homogeneous (also in time) relaxation. However, it should be noted that the relative deformation $\Delta L/L$ over which we compute the elastic modulus is extremely small, for a typical MD simulation (see ordinate axis in Fig. 6), therefore the amplitude of force fluctuations appears larger on this scale. The Young's modulus can be extracted from the linear fit of the data in Fig. 6, as:

$$(1) \qquad \frac{f}{A} = Y \left( \frac{\Delta L}{L} \right)$$

with $A$ the cross section of the nanowire (assumed cylindrical), defined by the previously estimated value of average radius $R$=0.95 nm. Our best estimate is $Y$=1.8±0.5 GPa, a quite large value compared to the Young's modulus of B-DNA, which lies rather in the range of 0.35 GPa (39). For the sake of comparison, the inset of Fig. 6 shows the results of a run with spring constant of 7 kcal mol$^{-1}$ Å$^{-2}$ and pulling velocity of  5 ms$^{-1}$: for this case we obtain a Young's modulus about half the previously estimated value. Such a softening is a clear signature of the too rigid  spring constant, which gives rise also to wider force oscillations.

   We probed the nanowire toughness in stretching well beyond the linear elastic regime, without detecting any signs of mechanical instability up to forces of the order of 1000 pN and more. This would correspond to a tensile strength exceeding 300 MPa, notably as good as a mild steel or aluminum alloy wire. Also in this respect, the *i–motif* appears to differ substantially from B–DNA, which is known to undergo a kind of structural, or "melting" transition above a tensile force of ~65 pN (39) (a fact that also complicates the direct measurement of $Y$).

### *Bending stiffness*

Experimental measurements of the bending rigidity of polymers are based on variants of either one of two, quite general approaches: (i) exploit the dependence of thermal fluctuations on the stiffness, or (ii) measure the force needed to actively bend the polymer. In the first approach, the fluctuations of free filaments are monitored as a function of wavelength by light microscopy (also called "flicker spectroscopy" (40)). The second approach is typically implemented by measuring the force needed to push, e.g. by an AFM tip, a single filament deposited on a nanoscale patterned surface (41–43), in analogy with a macroscopic three–point bending measurement.

Both such experimental techniques can effectively be mimicked by MD simulation (44). The second one is but a variant of the tensile experiment described in the previous section, the force being applied at the center of the polymer while the two extremes are held fixed. The free fluctuation, however, is more complicate to observe on the MD time scale, since it requires that either (a) the contour length of the simulated polymer be much longer than its persistence length, or (b) the simulation time be long enough with respect to the longest-wavelength fluctuation relaxation time, this being in turn inversely proportional to the lowest frequency and, therefore, directly proportional to the contour length.

In fact, we tried to observe the free fluctuation of our longest *i–motif* nanowire (230 nm) during quite long, constant–{NVE} MD simulations. However, even for simulation times exceeding 10 ns, the nanowire remained practically straight, suggesting that its persistence length should be at least in the few–hundred nm range.

Therefore, we had to resort to the direct deformation by a simulated three-point bending experiment. Instead of the SMD, we found it more practical to use the "umbrella sampling" feature of NAMD (23), to smoothly move the *i–motif* along the bending trajectory, while recording the force at the midpoint. We fixed both the prime ends of the *i–motif* with contour length $L$=28.8 nm, by using a harmonic constraint on the four terminal C atoms. We then selected the mid point (the 24[th] nucleotide of each chain) and with the "umbrella" potential we pulled at constant velocity this set of four nucleotides along the x–axis (perpendicular to the main z–axis of the nanowire), up to a distance of about $\delta$=5 nm (see Fig. 7, top panel). We had to adjust the water box size, since the lateral extent of the bent structure exceeded the standard box dimensions, thereby starting to experience interaction with its periodic image.

While pulling the central nucleotide, at every step of 1 nm we fixed all three extremes of the *i–motif* (end-groups and pull-group) using harmonic constraints, and equilibrated the bent structure under constant–{NPT} for 5 ns, at $P$=1 atm and $T$=300 K. After such equilibration step, the bending deformation also adds a variable amount of stretching, which amounts to a maximum of ~2 nm for the maximum bending of 5 nm. Therefore, to get back to a contour length as close as possible to that of the initial structure and get rid of excess deformation energy, we performed a cycle of end-to-end compression for about 1 ns, during which the length was adjusted until the residual extension/compression was within ±0.3%. This compression step was followed by another constant–{NPT} equilibration cycle of about 2 ns. In Figure 7 we plot the force vs. displacement curve measured at the central (pull) group, together with a scheme showing the geometry of the simulated bending experiment.

If the *i–motif* nanowires were homogeneous and isotropic, the bending stiffness would be easily obtained from the Young's modulus as $B = Y \cdot I$, with $I$ the second moment of the transverse cross section, which, for a homogeneous cylinder of radius $R$, would be $I = \frac{\pi}{4}R^4$. From this simple relationship we get an estimate of $B$=1.7 · $10^{-27}$ N m$^2$.

A better estimate can be obtained from the calculation of the displacement $\delta$ at the midpoint produced by a point load $f$, as the derivative of the elastic energy with respect to the load (beam theory, Castigliano's theorem (45)):

$$(2) \qquad \delta = \frac{\partial}{\partial f}\left(\int_{-L/2}^{L/2} \frac{M^2(l)}{2B} dl\right) = \frac{f\,L^3}{48B}$$

$l$ being a variable spanning the contour length $L$, and $M(l) = \frac{1}{2} f \cdot l$ the bending moment, assumed to be linear along the contour length for a straight bar. From our force plot as a function of $\delta$ in Fig. 7, we can therefore get the following estimate for the bending stiffness:

$$(3) \qquad B = \frac{\alpha L^3}{48}$$

with $\alpha$=0.062 N m$^{-1}$ the linear slope of the $f(d)$ curve in Fig. 7 (dashed line), resulting in $B$=2.6 $\pm$ 0.4 $\cdot$ 10$^{-26}$ N m$^2$, considerably larger than the simpler estimate above.

We note that Eq. 2 neglects any shear effects in the wire, which, at the molecular scale, would manifest as small rearrangements of the bases about the backbones of the four intercalated chains. Moreover, using a constant $B$ in Eq. 2 implies the same approximation above for the second moment $I$. Also, some residual elastic energy from stretching and/or compression remains in excess after each equilibration. However, all such uncertainties are within the (conservatively large) quoted error bar.

### Persistence length

In our first attempts at studying the free fluctuation of the *i–motif* nanowire we wanted to use the fluctuation data also to extract the persistence length $\lambda_p$. As said above, such an attempt was frustrated by the quite high bending rigidity observed. It is worth noting that, from a statistical mechanics point of view, $\lambda_p$ has the meaning of a correlation length, expressing the distance over which the expectation value of the tangent between any two points, along the contour length $L$ of the polymer, becomes exponentially uncorrelated. On the other hand, the *mechanical* significance of $\lambda_p$ is rather that of a length scale: if $L \gg \lambda_p$ the polymer will appear easily folded and flexible; conversely, if $L \ll \lambda_p$ the polymer will appear very stiff.

Despite the difficulty of directly observing the fluctuations, an estimate of the $\lambda_p$ can also be provided from the above values of the bending stiffness. In fact, thinking of the thermal fluctuations as a stochastic force acting on the free polymer, $\lambda_p$ can also be defined as the length over which the energy of thermal fluctuations is comparable to the elastic energy required to bend a length $\lambda_p$ of the polymer:

$$(4) \qquad \lambda_p = \frac{B}{k_B T}$$

This definition gives an estimate of $0.4 < \lambda_p < 6$ mm, according to the two extreme values of $B$ deduced in the previous section.

## DISCUSSION

In this work we obtained the equilibrium molecular structures of *i–motif* DNA nanowires with a repeated $n[(TC_5)_4]$ sequence, by means of molecular dynamics (MD) simulations with the *Amber95* force field parametrization. In agreement with the experimental evidence accumulated up to date, such nanowires are assembled by intercalation of the four identical single DNA strands, arranged to give an interplanar distance between base pairs of 0.32 nm, close to the 0.34 nm spacing of B–DNA. We could confirm the relevant role of the extra proton, resonantly exchanged between co-planar cytosine pairs. In fact, upon MD simulation at $T$=300 K and $P$=1 atm, the non–protonated structure was found to be dynamically unstable, while the protonated one preserved its structural integrity over the time scale accessible to large–scale MD. Also in good agreement with experimental findings, we observed an alternance along the nanowire of the structure of T•T tetrads. These can be found alternately arranged inwards to the cylindrical axis and paired by hydrogen bonds, or freely floating outwards the main axis. Interestingly, it has been suggested that such an alternating behavior of the thymine pairs, which do not appear to accept any extra protons to increase their stability, could be at the basis of a biological switching mechanism (24).

By looking at the ensemble of mechanical properties, the Young's modulus under uniaxial deformation, the bending stiffness and persistence length under three-point bending deformation, we can make a meaningful comparison between the *i–motif* and other, more common biopolymers (see Table 1). First of all, we note that the *i–motif* appears to belong with the class of structural polymers (F-actin, microtubules, keratin, etc.), as far as its Young's modulus, while it appears nearly metallic with its tensile strength estimated at >300 MPa. With its $Y$ in the GPa range, the *i–motif* is at least one order of magnitude stronger than the DNA strands by which it is made up. On the other hand, if we look at its transverse flexibility, as measured by the apparent bending stiffness, the *i–motif* seems closer to DNA, and 1 to 3 orders of magnitudes more flexible than F-actin or microtubules. The reason is that the bending stiffness is mostly determined by the cross section squared, appearing in the second moment of the cross section, *I*. Compared to F–actin, the *i–motif* might look similar in terms of Young's modulus, and quite close in terms of bending. However, the Van–der– Waals bonded actin monomers are very easily broken apart, compared to the stronger hydrogen bond network of the C•C+ tetrads in the *i–motif*. Depending on the physiological conditions, F–actin breaks at a force of 230–260 pN (48), which, given its diameter of 5 nm, corresponds to a tensile strength of about 10 MPa, much smaller than the lower limit of 300 MPa we observed for the *i–motif*. Such a peculiar combination, of a very high tensile modulus and high toughness, and a relatively small bending modulus, makes the *i–motif* a rather peculiar nanowire, from the mechanical point of view: almost *inextensible* and *unbreakable*, but quite *flexible*, somewhat like a nanoscale steel chain. Of course, one should not confuse the tensile strength with the unfolding force, which is reportedly much smaller (49).

By looking again at Table 1, we wish also to underscore the comparison between *i–motif*, B–DNA, chromatin fiber, and the mitotic chromosome. The tensile stiffness of these objects rapidly decreases, from GPa to kPa, the chromatin and chromosome being extremely soft structures very easily coming apart under quite a small force (per unit cross section area). It might not be without implications that the small portions of *i–motif*, possibly contained in the chromosome, should behave like "hot spots" (for example during the transcription phase, when histone–binding proteins exert strong mechanical forces on the chromatin) concentrating the stress around the much tougher and stiffer sites, but still easily flexible, corresponding to the C–rich fragments of the sequence.

Indeed, it is worth noting that the repeated C–rich sequences of centromeric and telomeric regions have been found to fold into an intramolecular *i–motif* (3–5). Such an occurrence could impart an unusual toughness and rigidity to some regions of the genome. One might speculate about the genetic meaning, or function, of such a possibility. Just for the sake of argument, telomere length in white blood cells has been repeatedly observed to have an inverse correlation with blood pressure (50), it is directly related with loss of elasticity of arterial wall (51), and its shortening is increasingly accepted as a predictive biomarker for cardiovascular disease (52). It is highly speculative to ask whether such an effect in leucocytes could also have a mechanical component, stemming from a higher rigidity of telomere regions, ultimately linked to the possible presence of *i–motif* structures. If at least partly chromatin integrity could be associated with the presence of long telomeres, one may wonder whether the age–related, progressive shortening of telomeres could also imply the absence of tougher *i–motifs* segments, someway contributing to cell aging via easier DNA degradation.

## CONCLUSIONS

Besides their potential genetic significance, yet to be fully understood, *i–motif* nanowires with their peculiar mechanical properties could also represent very good candidates for biomimetic nanoscale templates. As recalled in the Introduction, bio–inspired nanowires are high on the wish–list of nanotechnology, for their promising potential of "easily" obtaining tailored and cheap self–assembled, nm–scale structures, onto which other nanometric objects with pre–designed functional properties (electric, electronic, magnetic, optical, etc.) could be assembled with a high degree of precision, e.g. via selective ligand–ligand interactions at well defined sites of the bio–nanowire. The self–assembled scaffold of *i–motif* tetramer can be cheaply arranged by design from simple DNA strands, into long–time stable nanowires (8) and, as we showed here, with highly peculiar mechanical properties. DNA is a common target for antiviral, antibiotic, anticancer drugs, capable of intercalating in the double–helix structure, or adsorbing at the major/minor grooves. Such a high ligand selectivity is likely to persist also in the *i–motif* parent structure, thus making for identification and addressing of preferential target sites: an ideal feature for nanostructure self–assembly. As a highly flexibile, but very stiff and very tough nanostructure at the same time, the *i–motif* could ideally serve to support structural and functional components in complex nanoscale devices.

## Acknowledgements

Computing grants from French Supercomputing Center IDRIS, and from CEA–TGCC (in the frame of the PRACE 2010–030294 Project to F.C. and R.B.) are acknowledged. R.P.S. gratefully thanks the President of the University Lille I for a collaborative, three-year PhD grant. We gladly acknowledge useful discussions with Dominique Collard (LIMMS Cnrs, Tokyo) who firstly suggested us to investigate the *i–motif* structure.

**TABLES**

Table I – Comparison of some mechanical properties for various biopolymers.

| | Young's modulus (Gpa) | Bending stiffness (N m$^2$) | Persistence length (mm) |
|---|---|---|---|
| $n[(TC_5)_4]$ *i-motif* [1] | $1.8 \pm 0.5$ | $(1.9 \pm 0.5) \cdot 10^{-27}$ to $(2.6 \pm 2) \cdot 10^{-26}$ | 0.4 to 6 |
| F-actin [2] | 2.6 | $7.3 \cdot 10^{-26}$ | 17.7 |
| Microtubules (taxol stabilized [2]) | 1.2 | $2.2 \cdot 10^{-23}$ | 5200 |
| Microtubules (bundles from pillar cells [3]) | 2 | $7 \cdot 10^{-23}$ | |
| Wool keratin [4] | 4 | | |
| Single vimentin intermediate filament [5] | 0.9 to 2.4 | $4 \cdot 10^{-27}$ | ~1 |
| Elastin [4] | $6 \cdot 10^{-4}$ | | 4 to $6 \cdot 10^{-4}$ |
| B-DNA [6] | 0.35 | $2 \cdot 10^{-28}$ | 0.05 |
| Chromatin (chicken erythrocite [7]) | $7 \cdot 10^{-6}$ | | 0.03 |
| Chromosome (mitotic phase [7]) | $5 \cdot 10^{-7}$ | | |

[1] this work ; [2] Ref. 40 ; [3] Ref. 42 ; [4] Ref. 46 ; [5] Ref. 43 ; [6] Ref. 39 ; [7] Ref. 47

**FIGURE CAPTIONS**

Figure 1. (a) Side view of the $n[d(TC_5)]$ basic dimer in stick representation, with the two chains colored differently, as obtained from the crystallography–resolved structure and duplicated $n$ times. (b) Side view of the $n[(TC_5)_4]$ intercalated tetramer, used as starting configuration for MD simulations. The intercalated yellow–cyan chains are identical to the green–magenta pair, rotated by ~50° and shifted by 0.31 nm about the z–axis. (c) Top view of the intercalated tetramer in a different graphical representation, showing the double–grooved secondary structure.

Figure 2. Definition of the various distances of the nucleotides in the four chains composing the *i–motif* tetramer. Average values obtained after the relaxation–equilibration molecular dynamics cycle described in the text. (a) IP=interplanar (or base stacking) distance (see also (b)); IC=intercalation distance (see also (c)); D=diameter; MiG=minor groove width; MaG=major groove width.

Figure 3. Comparison of the structure of the protonated vs. non-protonated *i–motif* tetramer after 5 ns of molecular dynamics at T=300 K. The central (b) structure in blue is the initial configuration, identical for both cases; (a) the protonated *i–motif* tetramer; (c) the non–protonated tetramer structure; (d), the two structures superimposed.

Figure 4. Comparison of the structure of the protonated vs. non–protonated *i–motif* tetramer (snapshot of one chain from each tetramer) after 5 ns of molecular dynamics at constant–{NVT}. The blue chain is the first time–step, identical for both simulations; (a) the green chain is the protonated *i–motif*; (b) the red chain is the non-protonated tetramer with the same nucleotide composition; (c) protonated (green) and non–protonated (red) chains superimposed.

Figure 5. (a) Radial density of water molecules around the *i-motif* nanowire, expressed as a fraction of the normal density at STP. The zero of the abscissa corresponds to the central z–axis. The arrow indicates the outer *i–motif* surface (approximate radial position of the P backbone). Symbols are averages over axial slices of 6 nm, taken at different positions along the axis. (b) Radial molar density of $Na^+$ (blue) and $Cl^-$ ions (red) around the *i-motif*, averaged over the entire nanowire length. (c) Cross–section (9x9 $nm^2$) of the water molecules (dots), $Na^+$ and $Cl^-$ ions (blue and red squares), in a thickness of 6 nm along the z–axis in the central portion of the nanowire, projected in the xy-plane normal to the cylinder axis. The thick circle indicates the average position of the *i–motif* major groove (green) and minor groove (orange).

Figure 6. Force-displacement plot for uniaxial stretching deformation of the *i-motif* tetramer, for a constant pulling velocity of 1 $ms^{-1}$ and a spacer spring constant of 1 kcal $mol^{-1}$ $A^{-2}$. Top: scheme of the bending simulation. The thick dashed line is the linear fit to the average mechanical response. The inset shows a simulation with pulling velocity of 5 $ms^{-1}$ and spring constant of 7 kcal $mol^{-1}$ $A^{-2}$.

Figure 7. Force-displacement plot for bending deformation of the *i-motif* tetramer. Top: scheme of the bending simulation. The red continuous line is a guide to the eye (quadratic fit); the thick dashed blue line is the best linear fit.

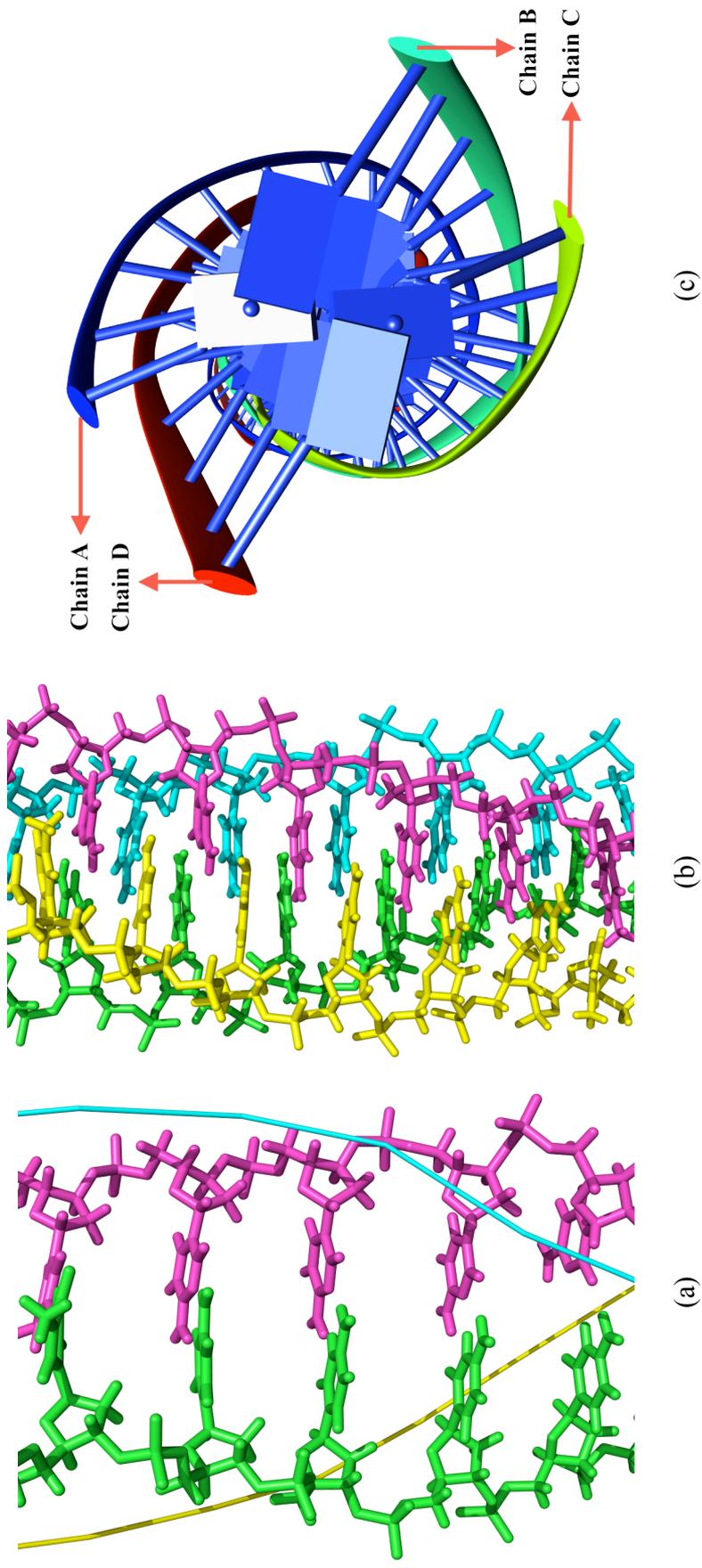

Figure 1

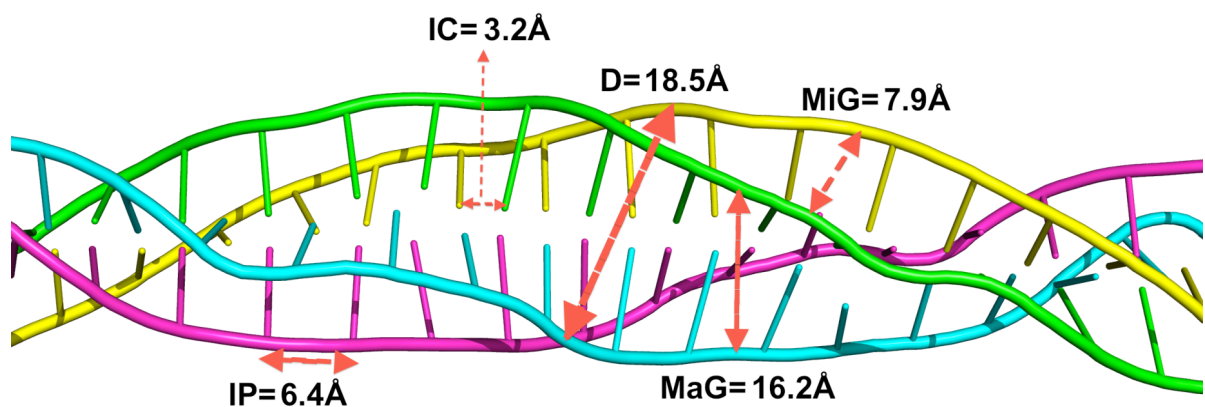

(a)

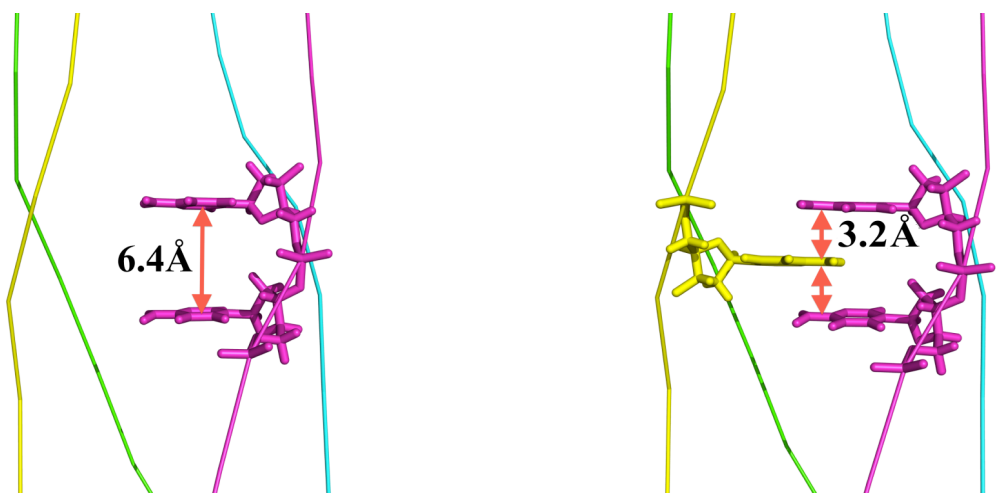

(b)                                                    (c)

Figure 2

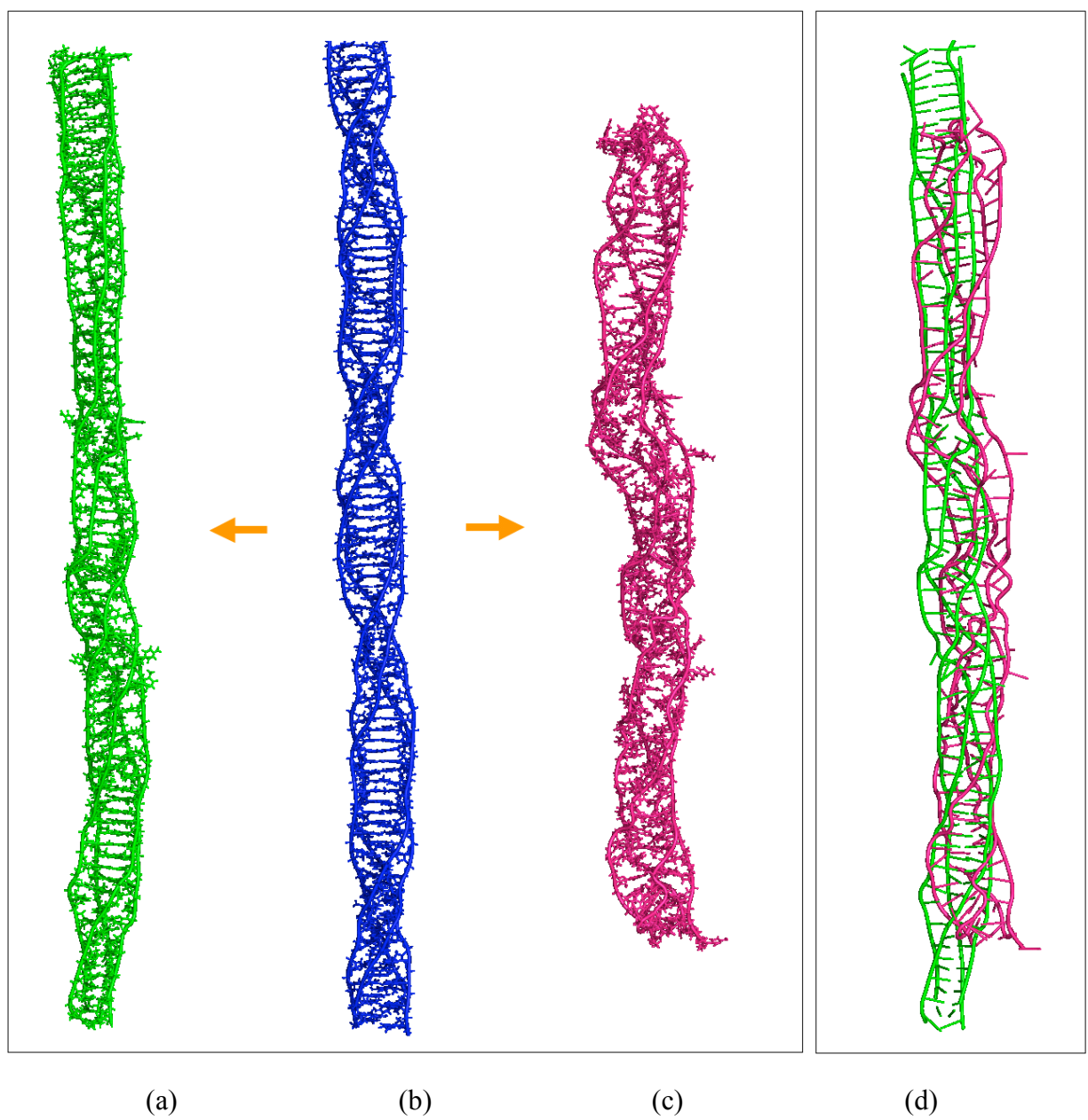

(a)  (b)  (c)  (d)

Figure 3

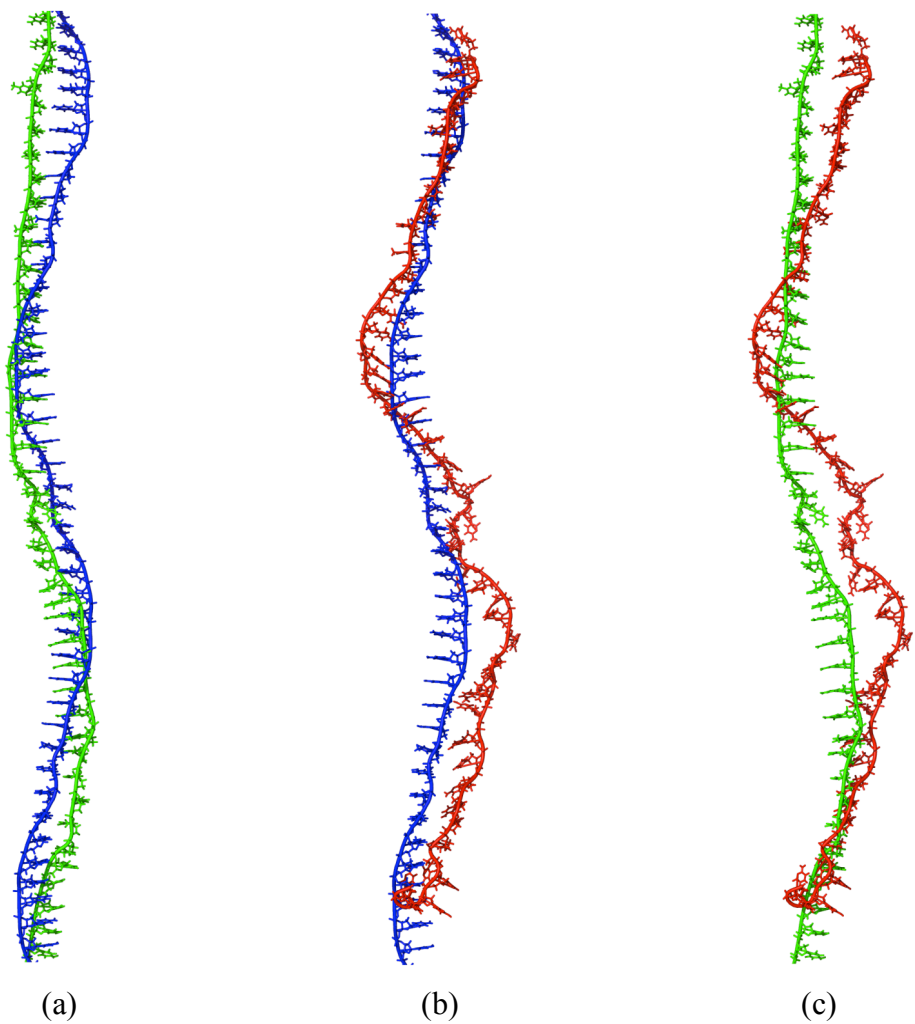

(a)  (b)  (c)

Figure 4

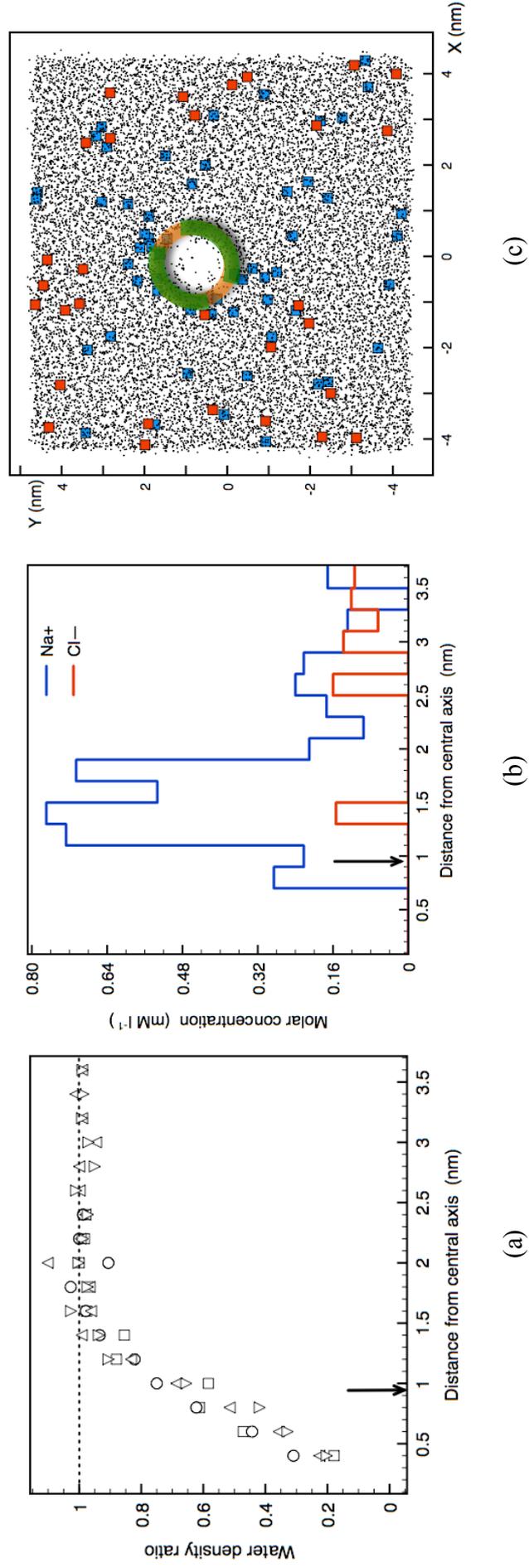

Figure 5

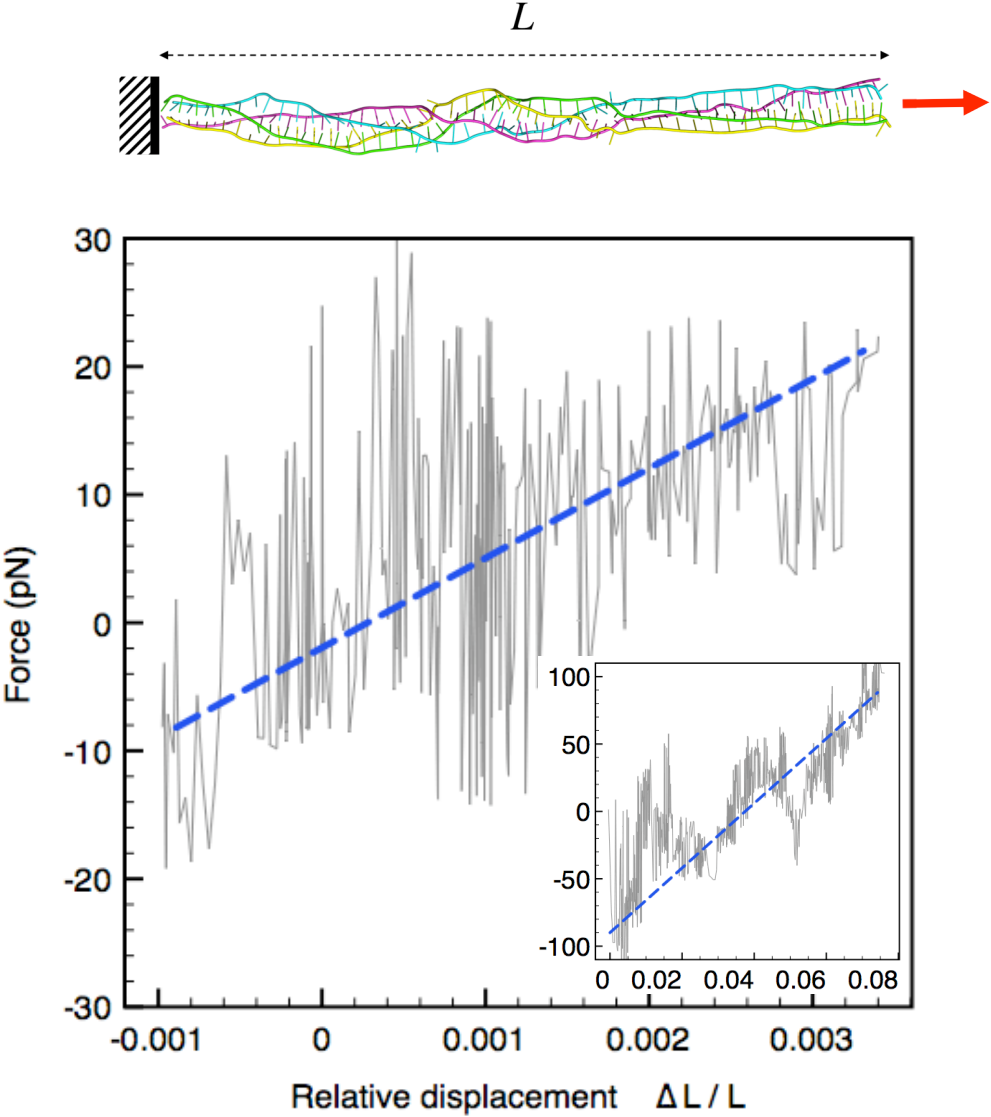

Figure 6

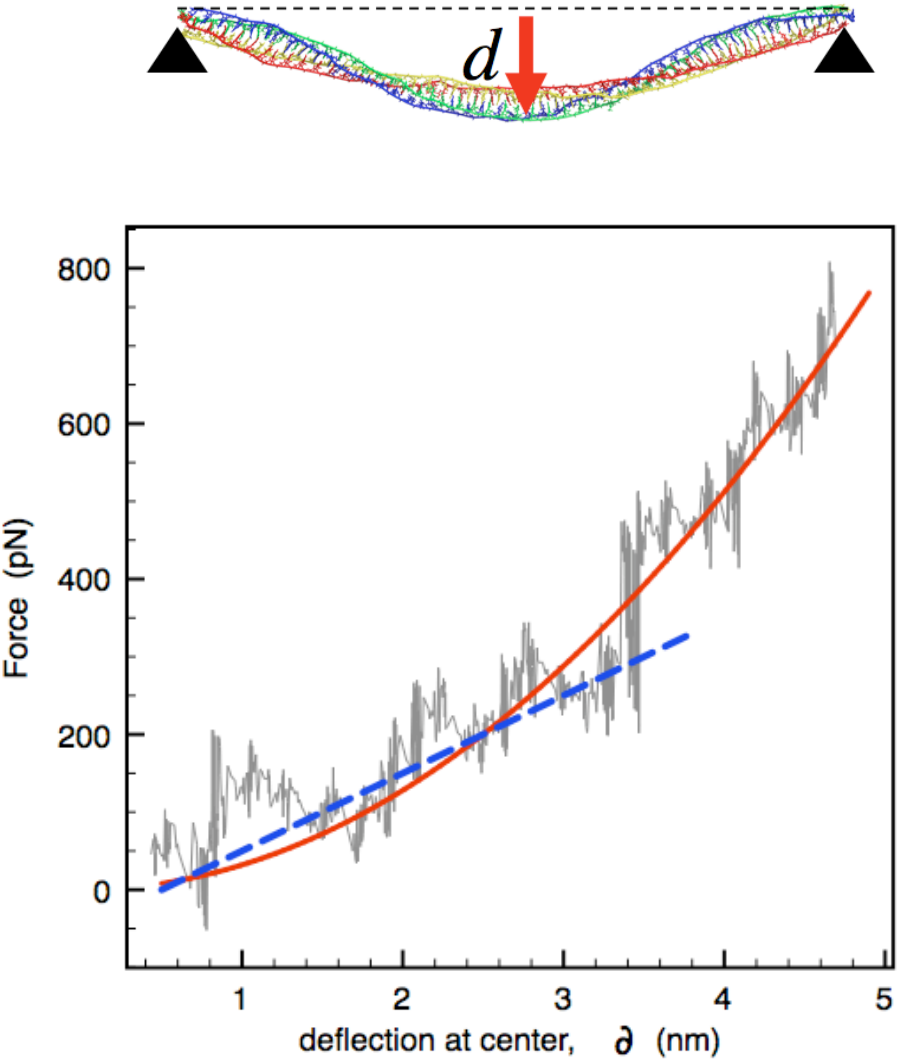

Figure 7